\documentclass[twocolumn,showpacs,showkeys,preprintnumbers,amsmath,amssymb]{revtex4}

\usepackage{graphicx}
\usepackage{bm}

\begin{document}


\title{Guiding of cold atoms by a red-detuned laser beam of moderate power}

\author{B.T. Wolschrijn, R.A. Cornelussen, R.J.C. Spreeuw,
and H.B. van Linden van den Heuvell}

\affiliation{Van der Waals-Zeeman Institute, University of Amsterdam, \\
         Valckenierstraat 65, 1018 XE Amsterdam, the Netherlands\\
         e-mail: spreeuw@science.uva.nl}

\date{\today}

\begin{abstract}

We report measurements on the guiding of cold $^{87}$Rb atoms from a
magneto-optical trap by a continuous light beam over a vertical distance
of 6.5~mm. For moderate laser power ($<$85~mW) we are able to capture
around 40$\%$ of the cold atoms. We systematically study the guided
fraction as a function of laser power and detuning, and give an
analytical expression that agrees well with our results. Although the
guide is red-detuned, the optical scattering rate at this detuning
($\approx$70 GHz) is acceptably low. For lower detuning ($<$30~GHz) a
larger fraction was guided but radiation pressure starts to push the
atoms upward, effectively lowering the acceleration due to gravity.

\end{abstract}





\maketitle

\section{Introduction}

The development of laser cooling has stimulated a tremendous interest in the
confinement and manipulation of cold atoms. In particular the use of dipole
forces by far off resonance laser light has become a versatile tool for the
manipulation of atomic motion. In this paper we discuss the use of a
red-detuned guiding laser as a simple and effective method to transport cold
atoms, although we use only moderate laser power. We systematically study the
fraction of guided atoms by varying the main parameters laser power and
detuning. We also give an analytical expression that gives excellent agreement
with the results.

In our experiments so far a cold cloud of $^{87}$Rb atoms is collected in a
vapor cell magneto-optical trap (MOT) and dropped on an evanescent-wave (EW)
mirror \cite{CooHil82,BalLetOvc87,KasWeiChu90} after post-cooling in optical
molasses \cite{VoiWolCor01,VoiWolSpr00,WolVoiJan01}. However due to the finite
temperature of the cloud, it expands ballistically during the fall, resulting
in a decreased density at the surface. This density reduction of atoms falling
on the EW could be counteracted in several ways. For example, one could focus
the atomic cloud by means of a strong pulse of magnetic field gradient
\cite{MarGuiBos99}, or combined magnetic and optical pulses \cite{Bal97}.

Here we describe our method to guide the atoms by optical light fields,
confining them in the transverse direction. The advantage of an optical
guide is that it is a well controlled force, that can be applied very
locally. The disadvantage of this technique is that atoms potentially
scatter photons from the guide, which results in heating and loss of
atoms. To overcome this problem of scattering, some groups have used
blue-detuned hollow beams \cite{SonMilHil99,ManOvcGri98}, in which the
atoms spend most of their time in the dark inner region of the guide.
Such hollow beams cause a low optical scattering rate and can even
cool the atoms during reflections at the wall \cite{YinZhuWan98}.

The use of a red detuned laser guide has previously been described by several
authors. For example, it has been used to transport cold atoms from a
production region to a spatially separated trapping region
\cite{SzyDavAda99,DavAda00}. Adiabatic heating and cooling in converging and
diverging guiding beams was reported in Ref. \cite{PruMarHou99}. A beam
splitter for cold atoms has been demonstrated by crossing two red detuned
guides \cite{HouKadPru00}. These experiments have in common that they make use
of high optical power. Red detuned guiding has also been used to capture atoms
from a thermal vapor and guide them through hollow fibers \cite{RenMonVdo95}.

In this paper we investigate the simplest possible optical guide: a
red-detuned Gaussian laser beam of moderate power. The atoms act as
high-field seekers and are pulled toward the center of the guiding-beam
\cite{BjoFreAsh78}. Therefore, the photon scattering rate will be higher
than in a hollow guide. However, a Gaussian beam is straightforward to
implement, while the scattering rate may still remain at an acceptable
level for our purposes. We investigate the performance of the guide for
various optical parameters.

\section{Experimental procedure}
\label{expproc}

The experiment is performed in a glass vapor cell where we create a MOT
of $^{87}$Rb atoms, 6.5~mm above a BK7 glass prism. Above the prism
surface we create an evanescent wave (EW) that is used as an atomic
mirror \cite{VoiWolCor01,VoiWolSpr00}. The purpose of the guiding beam
is to confine the atoms while falling towards the EW mirror.

The Rb density in the cell is increased by heating a Rb reservoir. Differential
pumping of the cell ensures a constant density long enough to perform an
experiment, but it changes on a daily basis. Consequently also the width of the
cloud changes on a daily basis. The r.m.s. radius of the cloud, $\sigma$,
varies between $200~\mu$m and $700~\mu$m. The temperature of the cloud is
unaffected by the varying density and is $4.0\pm0.5~\mu$K after postcooling in
optical molasses.

\begin{figure}[t]
\includegraphics[width=70mm]{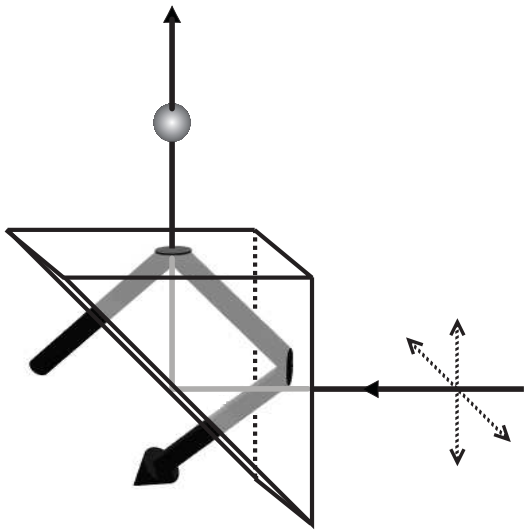}
  \caption[Experimental setup of the guiding laser-beam]
  {\it Experimental setup of the guiding laser. The guiding beam is pointed upwards by
  reflecting it at the hypotenuse of the prism.
  The beam must overlap with both the evanescent wave (depicted as the disk at the prism surface)
  and with the MOT (represented by the grey sphere).}
  \label{fig:guidesetup}
\end{figure}

The laser light for the guiding beam is obtained from a home-built
tapered amplifier system \cite{VoiSchSpr00}. The beam is spatially
filtered by sending it through a single-mode optical fiber and linearly
polarized by a polarizing beam splitter. The guiding laser was reflected
at the hypotenuse of the prism, such that it was pointed upward (see
Fig.\ref{fig:guidesetup}). For an optimal result the guiding beam must
overlap with the EW as well as with the MOT. To align the beam we
mounted some of the steering mirrors on translation stages. This allowed
us to translate the beam perpendicular to its propagation direction, and
to independently adjust its angle.

We measured the beam profile at various positions along the beam line. The
waist ($1/e^2$ radius) was found to be $550~\mu$m, located at the position of
the MOT. The divergence half-angle was $0.8$~mrad. We consider the beam waist
to be constant in the $6.5$~mm region between the MOT and the EW. The available
power was limited by the tapered-amplifier system and its transmission through
the fiber, which reached a maximum of $40$\%. We obtained a total power of
approximately $100$~mW, just before entering the vacuum cell. Because the
vacuum cell and the prism are not anti-reflection coated, the available power
for guiding is $85 \%$ of the power before the cell. Throughout this paper we
quote the power of the guiding beam \emph{inside} the cell, at the position of
the MOT. We were able to switch off the beam in $100~\mu$s by means of a
mechanical shutter, placed before the fiber. The guide was red detuned with
respect to the $D_{2}$-line, which was measured by a wavemeter (Burleigh,
WA-1500).

\begin{figure}[t]
\includegraphics[width=80mm]{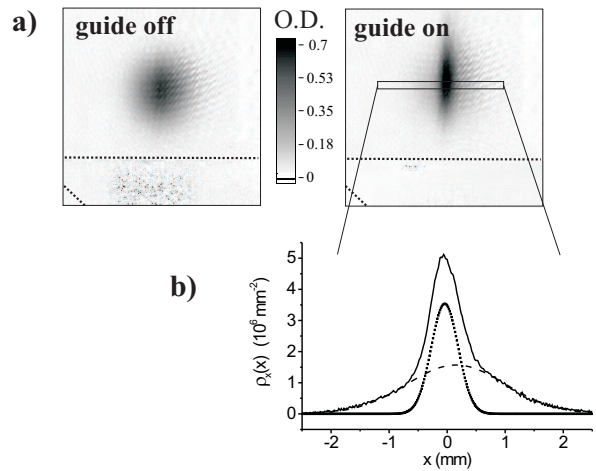}
  \caption[Image of guided atoms and fit]
  {\it {{\bf a)} Absorption images of falling atoms without (left image),
  and with the guiding beam present (right image). The field of view
  is $7.8~\times~7.8~$mm$^2$. The images are taken
  $28$ ms after releasing the atoms from the molasses, just before they arrive at the prism.
  The grey scale indicates the optical density (O.D.).
  The box indicates the part of the image which is averaged, in order to
  analyze the data.
   {\bf (b)} Atomic density along the horizontal direction.
   Solid curve: numerical average of $30$ horizontal rows.
   The dotted and dashed curve are the result of a fit, using two Gaussians,
   corresponding to the guided an unguided atomic distribution.
   In this example the guiding beam had a detuning of $-13 \times 10^{3} \Gamma$,
   and a power of $85$~mW.}
  \label{fig:guidefit}}
\end{figure}

The guiding beam was continuously on during the loading of the MOT, cooling in
molasses, and during the fall of the atoms toward the prism. For sufficiently
large detuning of the guide, we observed no effect on the MOT or molasses.
After cooling in optical molasses, the atoms fall down and a fraction of the
atoms is captured in the guiding beam. The atomic distribution is imaged on a
CCD camera by resonant absorption imaging at time $t_{p}$. Just before we probe
the atoms, the guide is switched off, in order not to blind the camera by stray
light from the prism. In Fig.~\ref{fig:guidefit}a we show two absorption images
of the atomic cloud, both after falling for $28$~ms (just before reaching the
prism), with and without the guiding beam present. It is clear that when the
guiding beam is on, the density is increased in the cigar-shaped area of the
cloud.

\section{Measurements of guided fraction}

In order to determine the guided fraction, we averaged the atomic density
distribution in the vertical direction over $30$ pixel rows, indicated in
Fig.~\ref{fig:guidefit}a by the thin box. The data are well fitted by a sum of
two Gaussians, corresponding to a guided and an unguided fraction
\cite{PruMarHou99}. The broad distribution represents the unguided atoms and
has a r.m.s radius of $850~\mu$m, which is the same as the radius of the atomic
cloud without the guiding beam present. The narrow distribution in
Fig.~\ref{fig:guidefit}b is the density distribution of the guided atoms. It
has a r.m.s. radius of $220~\mu$m, slightly smaller than the waist of the
guide. From the area of the fitted distributions we extracted the guided
fraction.

\begin{figure}[t]
\includegraphics[width=80mm]{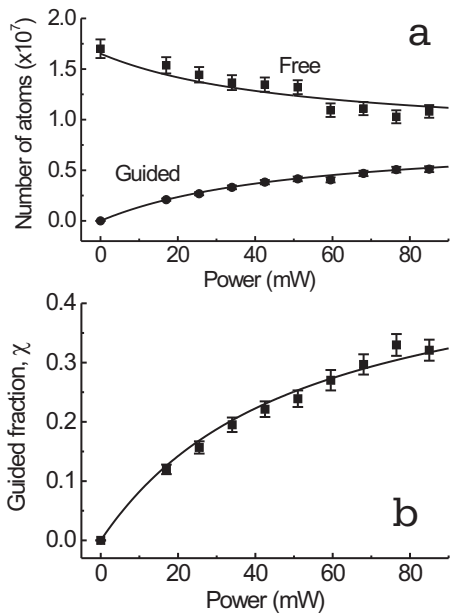}
\caption[Guided fraction vs guide-power] {\it {\bf (a)} Guided and unguided
number of atoms at $28$~ms, as a function of guiding power and {\bf (b)} the
fraction of guided atoms. The detuning and waist of the guiding beam are
$\delta = -12 \times 10^{3} \Gamma$, $w=550~\mu$m and the temperature of the
cloud is $4~\mu$K. The solid curve is the result of a fit according to
Eq.\ref{eq:guidefraction}, resulting in a r.m.s. radius $\sigma~=~650~\mu$m of
the molasses .} \label{fig:guidepower}
\end{figure}

We measured the fraction of atoms that were guided for $28$~ms as a
function of potential depth $|U_0|$. The depth can be changed by varying
either the power or the detuning of the guide. To calculate the
potential depth we take the light to be linearly polarized, and far off
resonance with respect to the $D_2$ line. The potential is then
approximately that of a two-level atom, multiplied by 2/3. This factor
is the sum of the squares of the Clebsch-Gordan coefficients, summed
over the excited-state hyperfine levels 5$P_{3/2}$, $F=0,1,2,3$. Note
that both hyperfine ground states experience approximately the same
potential. A Gaussian beam, with waist $w$, power $P$, and detuning
$\delta$ produces a potential depth at the center of the beam of
\begin{equation}
U_0\approx\frac{\hbar P\Gamma^2}{6\pi w^2I_0\delta},
\label{eq:potentialdepthguide}
\end{equation}
where $\Gamma=2\pi\times 6.1$~MHz is the natural linewidth of the $D_2$ line
and $I_0=1.6$~mW/cm$^2$ is the saturation intensity. Red detuning corresponds
to $\delta<0$.

First we measured the guided fraction as a function of guiding power, keeping
the detuning fixed at $\delta=-12 \times 10^{3} \Gamma$,
(Fig.~\ref{fig:guidepower}). Varying the power from $0$ to $85$~mW changes the
potential depth at the center of the beam from $0$ to $22~\mu$K. Increasing the
power of the guiding beam obviously results in a higher capture fraction of the
guide, since the potential becomes deeper. The largest guiding fraction we
measured at this detuning was 33~\%, at a power of $76$~mW. The error bars in
the graph are due to uncertainties in determining the number of atoms.

\begin{figure}[t]
\includegraphics[width=80mm]{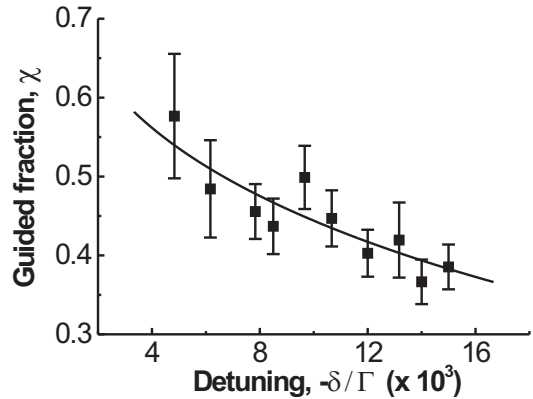}
\caption[Guided fraction vs guide-detuning] {\it Fraction of atoms that are
guided after $28$~ms, as a function of detuning of the guiding beam, $\delta$.
The squares are the experimentally obtained values and the solid curve is the
result of a fit by Eq.~\ref{eq:guidefraction}. The experimental parameters of
the guiding beam were: $w=550~\mu$m, $P = 85$~mW. The temperature of the cloud
was 4 $\mu$K. The fit resulted in a r.m.s. radius $\sigma = 540~\mu$m of the
molasses.} \label{fig:guidedetuning}
\end{figure}

We also varied the potential depth by changing the detuning, keeping the power
constant at $85$~mW. The guided fraction decreased only slightly when the
detuning was varied from $-4.9 \times 10^{3} \Gamma$ ($|U_0|/k_B=54~\mu$K) to
$-15 \times 10^{3} \Gamma$ ($|U_0|/k_B =18~\mu$K), see
Fig.~\ref{fig:guidedetuning}. As expected the fraction of guided atoms
decreased for larger negative values of the detuning, since this leads to
decreasing $|U_0|$. The range over which we could vary the detuning was limited
by the angle of the grating in the master laser of the tapered amplifier. The
error bars in Fig.~\ref{fig:guidedetuning} are much larger then the error bars
in Fig.~\ref{fig:guidepower}. This is because the demands on the stability of
the system are stronger for this set of measurements because it takes longer to
change the detuning than the power.

\section{Analysis and discussion}

To analyze the experimental data we first remark that the optical
potential describing the guide is cylindrically symmetric, and has no
influence on the atoms along the $z$ direction. The optical potential
describing the guide is written as
\begin{equation}
U(\rho) = U_{0}  e^{ - 2 \rho ^{2}/w^{2}} < 0 \label{eq:guidepotential}
\end{equation}
where $\rho$ is the transverse position in cylindrical
coordinates, $w$ is the laser waist and $U_{0}$ the maximum potential
depth (at the center of the guiding beam) which is negative due to the
sign of the detuning. We assume that the atomic cloud is described by
Gaussian distributions for the horizontal position and momentum
coordinates.
\begin{equation}
\Phi (\rho ,p) = \frac{e^{-p^{2}/2 Mk_{B}T}}{2 \pi M k_{B}T} \times \frac{e^{-\rho^{2}/2 \sigma ^{2} }}{2
\pi \sigma ^{2}}  \label{eq:guidemolasses}
\end{equation}
where $p=\sqrt{p_x^2+p_y^2}$ is the horizontal momentum,
$T$ is the temperature of the molasses, and $M$ the atomic mass. This
distribution is normalized as
\begin{equation}
\int_0^{\infty} \int_0^{\infty} \Phi (\rho ,p) \  2\pi \rho d \rho \ 2\pi p dp =1
\end{equation}
Atoms will be bound to the guide if the total of their kinetic and
potential energy is negative:
\begin{equation}
\frac{p^{2}}{2 M} + U(\rho) < 0 \label{eq:guidecondition}
\end{equation}

We denote the fraction of molasses atoms which are captured in the guide
as $\chi$. Combining equations \ref{eq:guidemolasses} and
\ref{eq:guidecondition} , the guided fraction $\chi$ is written as
\begin{equation}
\chi = 4 \pi^2 \int_{}^{}{}\int_{p^2 /2M \le -  U(\rho)}^{}{ \Phi (\rho ,p) \  \rho d\rho
 \ p dp }
\end{equation}
which can be evaluated analytically:
\begin{equation}
\chi = 1 - \frac{w^2}{4 \sigma^2}   \left(\frac{|U_0|}{k_B T}\right)
^{-\frac{w^2}{4 \sigma^2}}\Gamma\left(\frac{w^2}{4 \sigma^2},0,\frac{|U_0|}{k_B
T}\right) \label{eq:guidefraction}
\end{equation}
Here $\Gamma(a,b,c)$ is the generalized incomplete Gamma function \cite{math}.
Note that the above analysis is similar to that given in Ref.
\cite{PruMarHou99}. The analytical expression for the guiding fraction was not
given in that paper.

From this expression we see that the guided fraction is determined by only two
dimensionless parameters: the relative size of the guide compared to the
molasses size, $w/\sigma$, and the depth of the guide compared to the
temperature of the molasses, $U_0/k_B T$. The factor $4$ results from the
different definition of the waist of the guide $w$
(Eq.\ref{eq:guidepotential}), and the width of the molasses $\sigma$
(Eq.\ref{eq:guidemolasses}). As expected the guided fraction tends to $0$ when
either the potential depth or the width of the guiding beam goes to $0$. The
fraction increases monotonically when either $|U_0|$ or $w$ increases.

The guiding fraction according to Eq.~(\ref{eq:guidefraction}) is plotted in
Fig.~\ref{fig:guidepower} and Fig.~\ref{fig:guidedetuning} as the solid curve.
In the evaluation of Eq.~(\ref{eq:guidefraction}) the waist of the Gaussian
potential was set to the fixed value of $w=550~\mu$m. The maximum potential
depth at the center, $U_0$, was calculated using
Eq.~(\ref{eq:potentialdepthguide}). The temperature of the molasses $T$ was set
to $4.0~\mu$K. The r.m.s. width of the molasses was used as a fit parameter
since it was not independently measured and it shows large daily variations as
mentioned in section \ref{expproc}. Both experimental curves were well fitted
by the theoretical function Eq.~(\ref{eq:guidefraction}).

In the experiment where the potential depth is varied by varying the power of
the guiding beam, the fit resulted in a radius of the molasses of
$\sigma~=~650~\pm~7~\mu$m. The error bar represents the 95\% confidence
interval. The experiment where the potential depth is varied by changing the
detuning resulted in a fit of the radius of the molasses of $\sigma
= 540~\pm 8~\mu$m.

Not surprisingly the guiding fraction depends strongly on the size of the
molasses. When we compare Fig.~\ref{fig:guidepower} and
Fig.~\ref{fig:guidedetuning} we observe that for the same guide parameters,
P=85~mW and $\delta~=-12 \times 10^{3} \Gamma$, the increase of molasses size
reduces the guided fraction from $40\%$ to $32\%$. Around the experimental
parameters the guiding fraction theoretically depends on the size of the
molasses as $\partial\chi/\partial\sigma\approx-1.1~$mm$^{-1}$.

\begin{figure}[t]
\includegraphics[width=80mm]{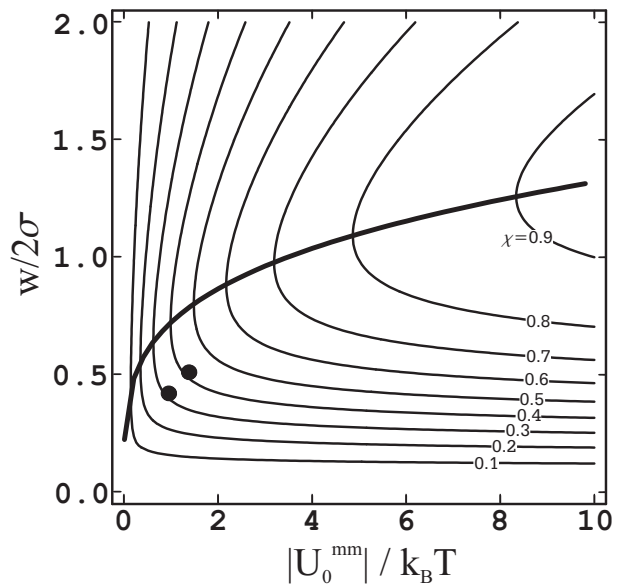}
\caption[optimumwaist versus Ur] {\it The thick curve shows the optimum
guide waist as a function of a "molasses-matched" potential depth
$|U_0^{mm}|$ in units of $k_B T$, which is proportional to the available
power and does not depend on the guide waist. When the experimental
parameters $\sigma$, $T$ and $P$ are known, the optimal guide waist $w$
can be read from this graph. The thin curves are curves of constant
guiding fraction $\chi$ (changing from $\chi=$0.1 for the left line to
0.9 for the right line). The points represent the experiment shown in
figure 3 (left) and 4 (right) with $P=85~$mW and
$\delta=-12\times10^3\Gamma$.} 
\label{fig:optimumwaist}
\end{figure}

Knowing how to characterize the guiding fraction, we can calculate the optimal
width of the guiding beam, given the available power and a chosen detuning for
which the scattering rate is at an acceptable level. When the waist of the
guide is small, the potential depth $|U_0|$ is large, but the spatial overlap
is not optimal. On the other hand, when the waist is very large, the spatial
overlap with the molasses is good, but fast atoms will escape from the low
potential. In Fig.~\ref{fig:optimumwaist} the thick line represents the optimum
guide waist as a function of the other experimental parameters, which are
summarized in a ``molasses-matched'' potential depth $|U_0^{mm}|$,
obtained by setting $w=2\sigma$ in equation (1), 
$U_0^{mm}=\hbar P \Gamma^2 / 24 \pi \sigma^2 I_0 \delta$. Note that this
parameter is proportional to the available power and does not depend on the
guide waist. The thin lines are lines of constant guiding fraction $\chi$.

Keeping the power at the maximum 85~mW and choosing the detuning at $\delta=
-12 \times 10^{3} \Gamma$, the optimum guide waists are determined to be
$920~\mu$m for the experiment shown in Fig.~\ref{fig:guidepower}
($\sigma=650~\mu$m) and $850~\mu$m for the experiment shown in
Fig.~\ref{fig:guidedetuning} ($\sigma=540~\mu$m). This leads to optimum guiding
fractions of 39\% and 48\% respectively. Our obtained results of $\chi=32\%$
and $\chi=40\%$ using a beam of $w=550~\mu$m, imply that improvement is
possible.

Our purpose of guiding the atomic cloud is to enhance the density of
atoms arriving at the surface of the prism, where the EW is located. For
$P=85$~mW at $\delta=-12\times 10^{3}~\Gamma$, the observed maximum
density of atoms inside the guide is $2\times 10^{9}$~cm$^{-3}$.
Comparing this to the value of a free, ballistically spreading cloud, we
see a density enhancement of $2.8$. The guiding fraction can be
increased to almost $60$\% by raising the depth of the guide to
55~$\mu$K. However such a large potential depth is obtained by tuning
the guide laser closer to resonance, at a detuning of
$-4.9\times~10^{3}~\Gamma$, where radiation pressure by the guide starts
to dominate the effect on the atoms, as will be shown in the next
section.

\section{Effect of photon scattering}

Up to now the photon scattering due to the guiding beam has not been
taken into account. In our experiments we make the potential deeper by
decreasing the detuning. Whereas the potential is linear in the detuning
($U \sim 1/\delta$), the scattering rate varies quadratically ($\gamma
\sim 1/\delta^2$). For a typical power of $85$ mW, and a detuning of
$-12~\times~10^3~\Gamma$, the photon scattering rate at the center of
the guide reaches a value of $250$~s$^{-1}$. During $28$~ms of falling,
$N=7$ photons are scattered on average. This corresponds to heating of
the atoms of $\Delta T=(N/3) T_{rec}= 0.8~\mu$K. This is small compared
to the initial temperature and we expect no influence on the guiding
properties. Indeed, we observed no increase of the width of the unguided
fraction when the power was increased.

\begin{figure}[t]
\includegraphics[width=80mm]{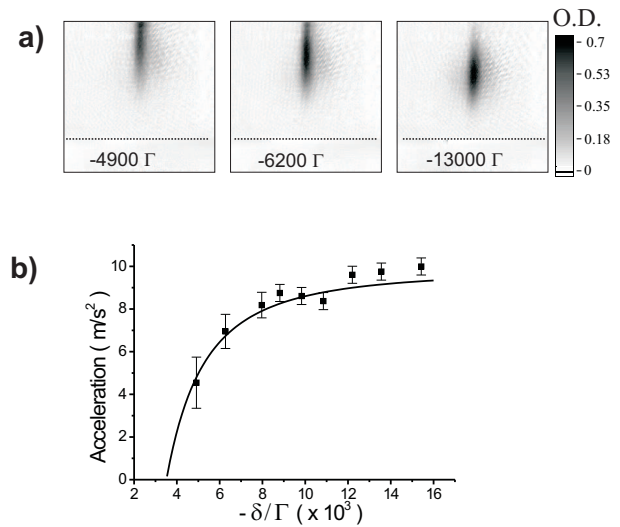}
  \caption[Low detuning pushes up the cloud]
  {\it
   {\bf a)} The atomic density profile after $28$~ms of guiding,
   for different detuning of the guiding beam. In each experiment the power
   was $85$~mW.
   {\bf b)} Measured acceleration of the guided fraction. For low detuning of the guide,
   the atoms are pushed upwards by radiation pressure.}
\label{fig:guidepushup}
 \end{figure}

The direct result of the high scattering rate at low detuning is
illustrated in the absorption images in Fig.~\ref{fig:guidepushup}a. All
three images are taken at $28$~ms after the molasses phase with the
guiding beam present, but with different detuning. For decreasing
detuning (closer to resonance), the scattering rate increases and
radiation-pressure pushes the atoms upward, along the direction of the
guiding beam. This results in an apparent gravity less than
$g=9.81$~m/s$^2$. For a detuning $\delta =-4.9 \times 10^{3} \Gamma$,
the radiation pressure is such that the guided atoms have fallen
$1.8$~mm instead of $3.8$~mm for free falling atoms, due to the
acceleration of only $4.5$~m/s$^2$.

We measured the atomic density profiles between $10$~ms and $36$~ms for
different values of the detuning ranging from $-4.8 \times 10^3 \Gamma$
to $-1.2 \times 10^4 \Gamma$, keeping the power at $85$~mW. For each
series we determined the vertical acceleration of the guided fraction
which is plotted in Fig.~\ref{fig:guidepushup}b. The error bars of the
experimental data are mainly due to the fact that the exact initial
position of the molasses is uncertain. This location is fitted to be
$0.63$~mm outside the field of view of the image. At low detuning, the
large errors are the consequence of the fact that a large fraction of
the guided cloud is outside the field of view.

The solid curve in Fig.~\ref{fig:guidepushup}b represents the calculated value
of the acceleration of the atoms in the guiding potential. To obtain this
curve, we calculated the exact dipole-potential and the maximum scattering rate
at the center of the trap, $\gamma _0$. The average scattering rate,
$\bar{\gamma}$ is less than this maximum value because the atoms oscillate in
the potential. We calculated the classical trajectories for 1000 atoms moving
in the dipole potential and time-averaged the intensity experienced by the
atoms. In our case the guide has a Gaussian shape, and under the experimental
conditions (T=4~$\mu$K) we obtain an average scattering rate $\bar{\gamma}
\approx 0.6~\gamma _0$. This ratio $\bar{\gamma}/\gamma_0\approx 0.6$ is
constant to within 5\% for the potential range of our experiments. The measured
apparent acceleration corresponds well with the theoretical value.

For values above $|\delta| = 12 \times 10^3 \Gamma$, we observed hardly
any influence of radiation pressure by the guiding beam. For such a
large detuning, the atoms scatter fewer than 5 photons during their
fall, and the guiding fraction is around $40$\%. This low scattering
rate implies that the atoms could be guided over larger vertical
distances. For lower detuning the atoms are pushed upward, resulting in
a delayed arrival time at the prism surface due to radiation pressure.
The increasing number of scattered photons results in heating of the
atomic cloud, visible as the elongation of the atomic cloud in the
vertical direction. For the lowest detuning, the width of the cloud is
almost 30\% larger compared to free falling atoms. During the time the
atoms were guided they have scattered on average 25 photons, resulting
in a temperature increase of 3 $\mu$K.

\section{Conclusions}

We showed the guiding of atoms from optical molasses by a continuous light beam
for 28~ms. For a total power of $85$~mW and detuning $\delta=-12 \times 10^3
\Gamma$, we guide $40\%$ of the atoms. Although the guide is red-detuned, the
optical scattering rate at this detuning is acceptably low. For lower detuning
a larger fraction was guided but radiation pressure starts to push the atoms
upward, resulting in an acceleration less than gravity. The results of the
measured guided fraction corresponded well with an analytical model. The
guiding fraction depends only on two dimensionless parameters: the ratio of the
widths of the guide and the molasses, and the ratio of the guide depth and
molasses temperature. Given the experimental parameters, an optimum value of
the guide waist can be determined. For a detuning of $\delta =-12 \times 10^3
\Gamma$, the density enhancement after 28~ms is 2.8, compared with
ballistically spreading atoms. This technique is easy to implement and can in
principle be used to guide atoms over larger vertical distances. The use of a
relatively small detuned laser beam of moderate power makes it easy to
implement a guiding beam into every experiment.

\section*{Acknowledgments}

This work is part of the research program of the ``Stichting voor
Fundamenteel Onderzoek van de Materie'' (FOM) which is financially
supported by the ``Nederlandse Organisatie voor Wetenschappelijk
Onderzoek'' (NWO). R.S. has been financially supported by the Royal
Netherlands Academy of Arts and Sciences.

\bibliographystyle{prsty}

\end{document}